\documentclass[a4paper]{article}
\usepackage{amsmath}
\usepackage{amsfonts}
\usepackage{amssymb}
\usepackage{graphicx}
\usepackage[labelfont=bf]{caption}
\usepackage[margin=1.8cm]{geometry}
\usepackage{epstopdf}
\usepackage{subcaption}
\usepackage{float}
\usepackage{pdfpages}

\makeatletter 
\renewcommand{\thefigure}{S\@arabic\c@figure}
\makeatother

\makeatletter 
\renewcommand{\thetable}{S\@arabic\c@table}
\makeatother

\makeatletter 
\renewcommand{\theequation}{S\@arabic\c@equation}
\makeatother

\begin{document}
\includepdf[pages=-]{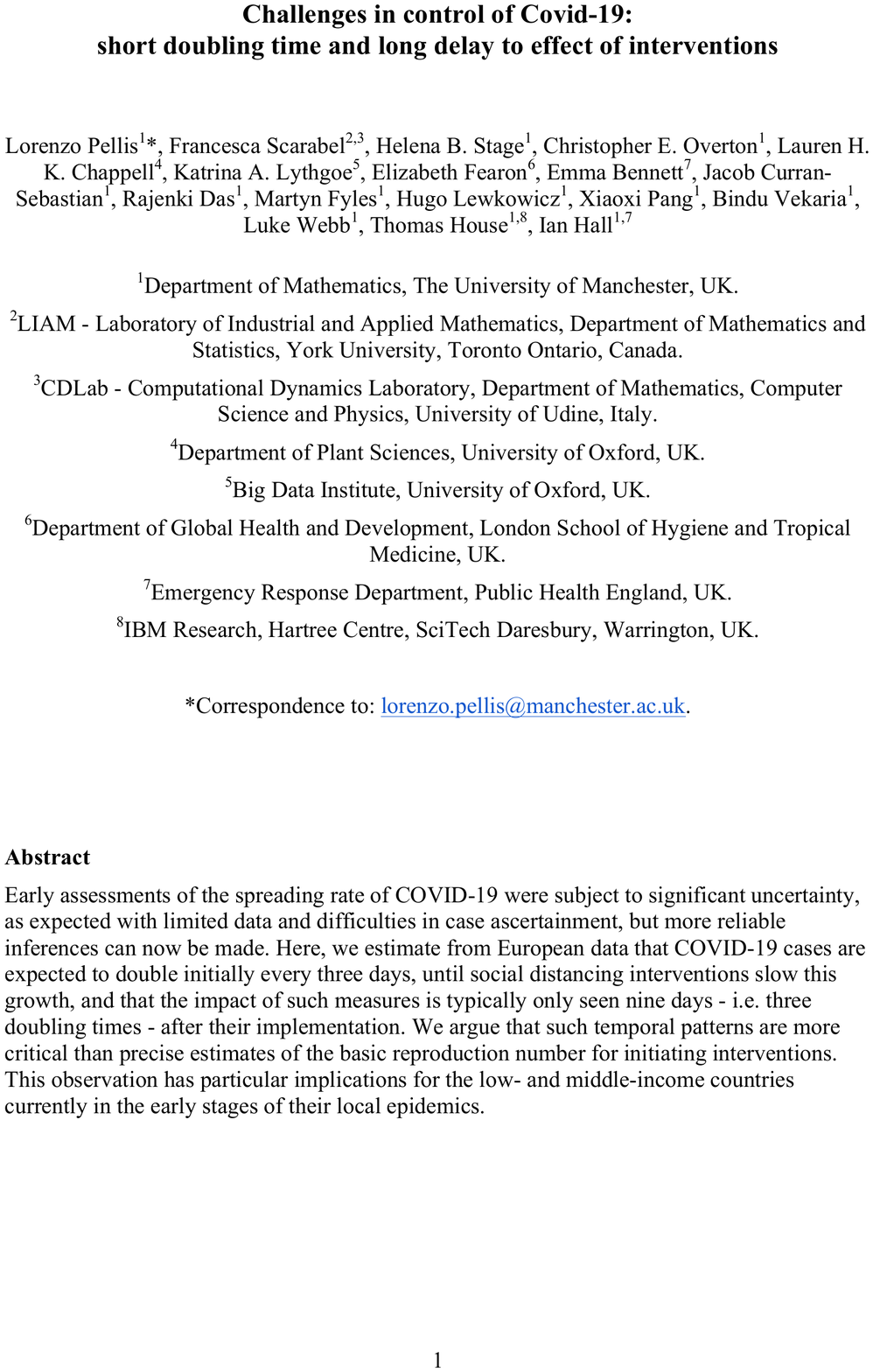}

\Large
 \begin{center}
Supplementary Materials for\\

\bigskip

Challenges in control of Covid-19: \\

short doubling time and long delay to effect of interventions

\hspace{20pt}

\small
Lorenzo Pellis$^{1,\ast}$, Francesca Scarabel$^{2,3}$, Helena B. Stage$^1$, Christopher E. Overton$^1$, Lauren H. K. Chappell$^4$, Katrina A. Lythgoe$^{5}$, Elizabeth Fearon$^6$, Emma Bennett$^7$, Jacob Curran-Sebastian$^1$, Rajenki Das$^1$, Martyn Fyles$^1$, Hugo Lewkowicz$^1$, Xiaoxi Pang$^1$, Bindu Vekaria$^1$, Luke Webb$^1$, Thomas House$^{1,8}$, Ian Hall$^{1,7}$\\

\hspace{10pt}

\small  
$^1$Department of Mathematics, The University of Manchester, Manchester, UK\\
$^2$LIAM - Laboratory of Industrial and Applied Mathematics, Department of Mathematics and Statistics, York University, Toronto Ontario, Canada\\
$^3$CDLab - Computational Dynamics Laboratory, Department of Mathematics, Computer Science and Physics, University of Udine, Italy\\
$^4$Department of Plant Sciences, University of Oxford, UK\\
$^5$Big Data Institute, University of Oxford, UK\\
$^6$Department of Global Health and Development, London School of Hygiene and Tropical Medicine, UK\\
$^7$Emergency Response Department, Public Health England, UK\\
$^8$IBM Research, Hartree Centre, SciTech Daresbury, Warrington, UK\\

\vspace*{10pt}
$^\ast$Correspondence to: lorenzo.pellis@manchester.ac.uk
\end{center}

\vspace{10pt}

\normalsize
\noindent\textbf{This PDF file includes:}\\
\indent Materials and Methods\\
\indent Supplementary Text\\
\indent Figs. S1 to S4\\
\indent Table S1\\

\noindent\textbf{Other Supplementary Materials for this manuscript include the following:}\\
\indent Data sources and the code used to carry out our data analysis can be found at:\\
\indent \texttt{https://github.com/thomasallanhouse/covid19-growth}
\newpage
\noindent\textbf{Materials and Methods}\\
\underline{Data Sources}\\
\indent We consider four sources for our epidemiological data: the WHO [\textit{28}], line-list data provided by Public Health England (PHE), line-list data from [\textit{33}] and the Italian Istituto Superiore di Sanit\`{a} [\textit{5}]. Of these data sets, three are publicly available. The line-list from PHE is unfortunately not publicly available. From these sources, we extract epidemiological data concerning case counts, incidence, hospitalisation, and delays between infection and symptom onset, and onset of symptoms and hospitalisation.\\
These data sources and the code used to carry out our data analysis can be found at:\\
\texttt{https://github.com/thomasallanhouse/covid19-growth}.\\

\noindent\textbf{Supplementary Text}\\
\underline{Fitting the growth rate}\\
\indent Typically, an infection spread from person to person will grow exponentially in the early phase of an epidemic. This exponential growth can be measured through the real time growth rate $r$ so that, loosely speaking, the prevalence of infection is 
\begin{equation}
    I(t) = I_0 {\rm e}^{rt} + \mathrm{noise.}
\end{equation}
A natural mathematical model to derive the estimate of $r$ is a Poisson family generalised linear model (GLM) with a log link. Given the over-dispersed noise inherent in both disease dynamics and surveillance data, a quasi-Poisson family is considered here. 

The growth rate $r$ is more intuitively reported as a doubling time (the time taken to double case numbers) and so $t_D=\ln(2) /r$. The log-linear analysis from a Poisson GLM defined formally below is restricted to datasets (or time windows) with clear exponential growth, or when additional explanatory variables, which are rarely available in real time, exist. 

To allow, in semi-parametric manner, time variation in growth rates we adapt a generalised additive model (GAM) where $I \propto {\rm e}^{s(t)}$ for some smoother $s(t)$. In particular, we use a quasi-Poisson family with canonical link and a thin-plate spline
as implemented in the R package \textit{mgcv} [\textit{45, 46}]. The instantaneous local growth rate is then the time derivative of the smoother $\dot{s}(t)$ and an instantaneous doubling time calculated as $t_D=\ln(2) /\dot{s}(t)$. Potential issues with the GAM approach include that extrapolation outside of the data range (and hence forecasting epidemic trend) is not sensible, and that there may be boundary effects from the choice of smoother. However, this approach has the major advantage that it allows for time-varying estimates of doubling time and thereby implicitly allows for missing explanatory information.

As well as the semi-parametric GAM approach, we take a parametric approach based on direct estimation of the exponential growth rate. This lacks the ability to capture time variation, but allows for extrapolation and epidemiological interpretation. To capture over-dispersion we use a quasi-Poisson family for the noise model. Explicitly, the Negative Binomial probability mass function (pmf) is
\begin{equation}
\mathrm{NB}(k|n,p) = \binom{k+n-1}{n-1} p^n (1-p)^k \text{ .}
\end{equation}
We will work in the parameterisation where the mean is $\mu$ and the variance is
$\theta \mu$, i.e.
\begin{equation}
p(\theta) = \frac{1}{\theta} \text{ ,} \qquad
n(\mu,\theta) = \frac{\mu}{\theta - 1} \text{ .}
\end{equation}
Let the number of new cases on day $t$ be $y(t)$. We assume that this is
generated by an exponentially growing mean,
\begin{equation}
\mathbb{E}[y(t)] = y_0 {\rm e}^{rt} = \mathrm{exp}(\ln(y_0) + rt) \text{ ,}
\end{equation}
which is then combined with the negative binomial pmf to give a likelihood
function for the observations over a set of times $\mathcal{T}$ of 
\begin{equation}
L(\mathbf{y}|y_0,r,\theta) = 
\prod_{t \in \mathcal{T}} \mathrm{NB}(y(t)|n(y_0 {\rm exp}(rt),\theta),p(\theta)) 
\text{ ,}
\end{equation}
where $\mathbf{y} = (y(t))_{t\in\mathcal{T}}$.  This can then be viewed as a
generalised linear model (GLM) with time as a continuous covariate, intercept
$\ln(y_0)$, slope $r$, exponential link function and negative binomial noise
model [\textit{47}]. We can perform inference through numerical maximum likelihood estimation
(MLE) and calculate confidence intervals using the Laplace approximation [\textit{48}].

\begin{figure}[h!]
    \centering
    \begin{subfigure}{.28\textwidth}
    \centering
    \includegraphics[width=0.95\textwidth]{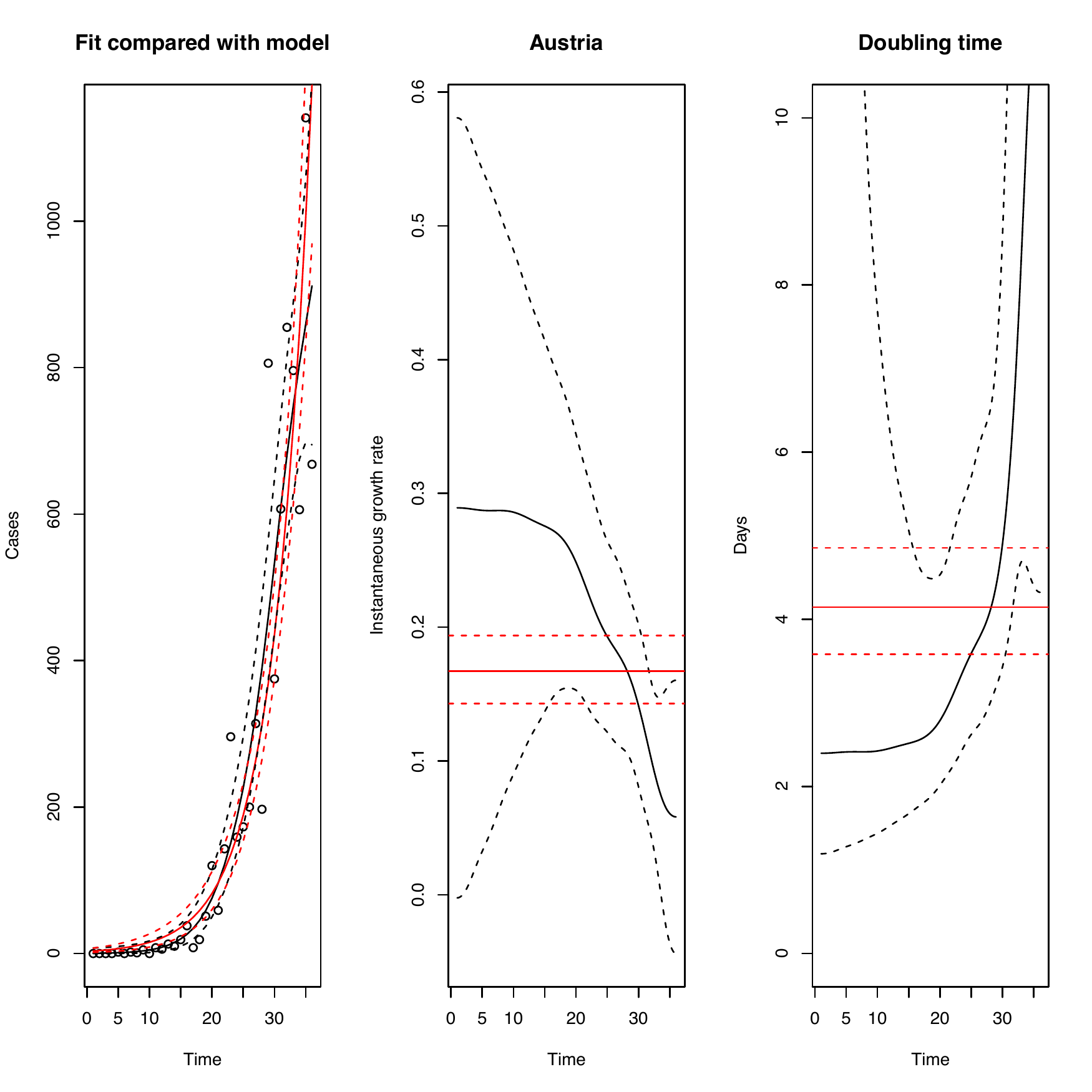}
\end{subfigure}%
\begin{subfigure}{.28\textwidth}
    \centering
    \includegraphics[width=0.95\textwidth]{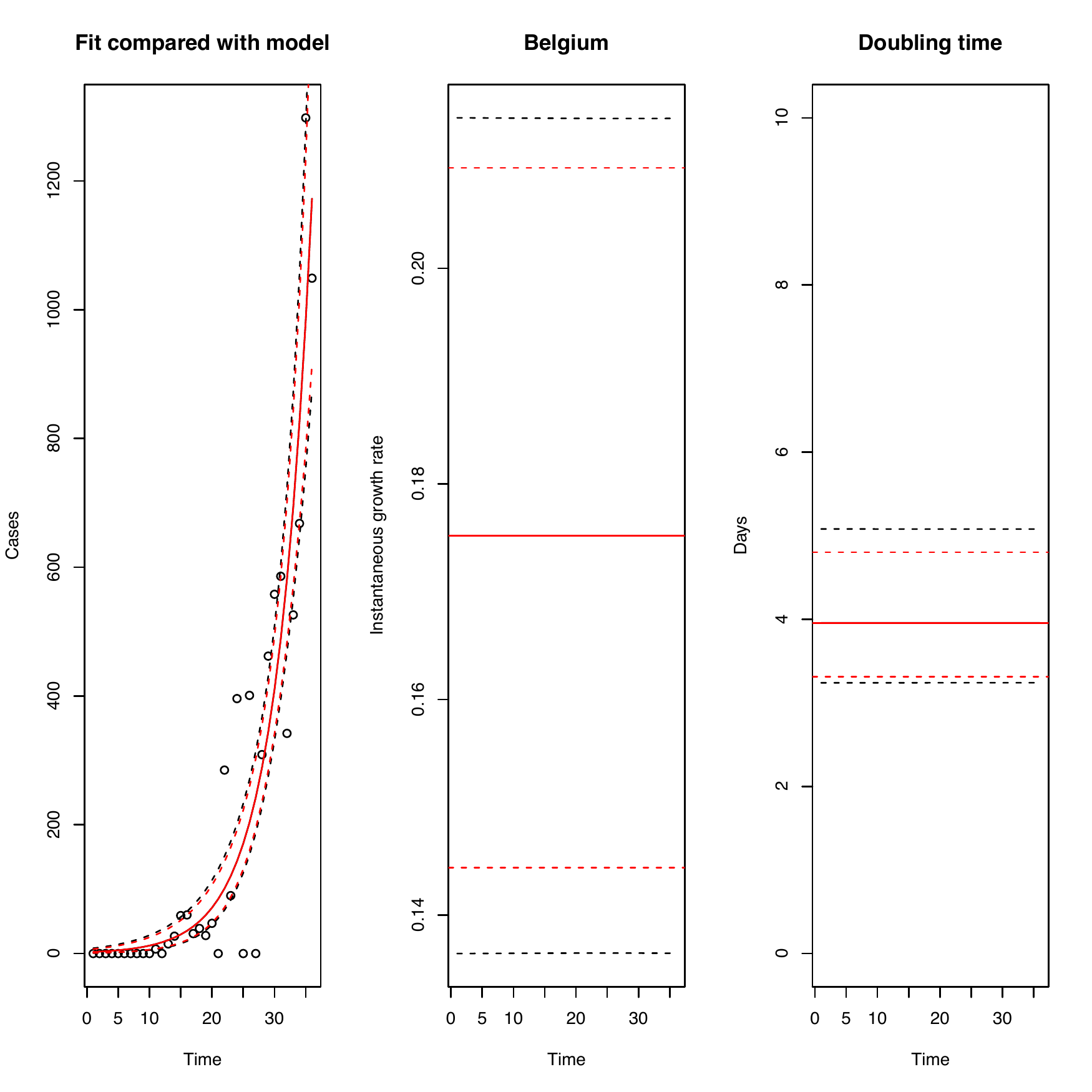}
\end{subfigure}%
\begin{subfigure}{.28\textwidth}
    \centering
    \includegraphics[width=0.95\textwidth]{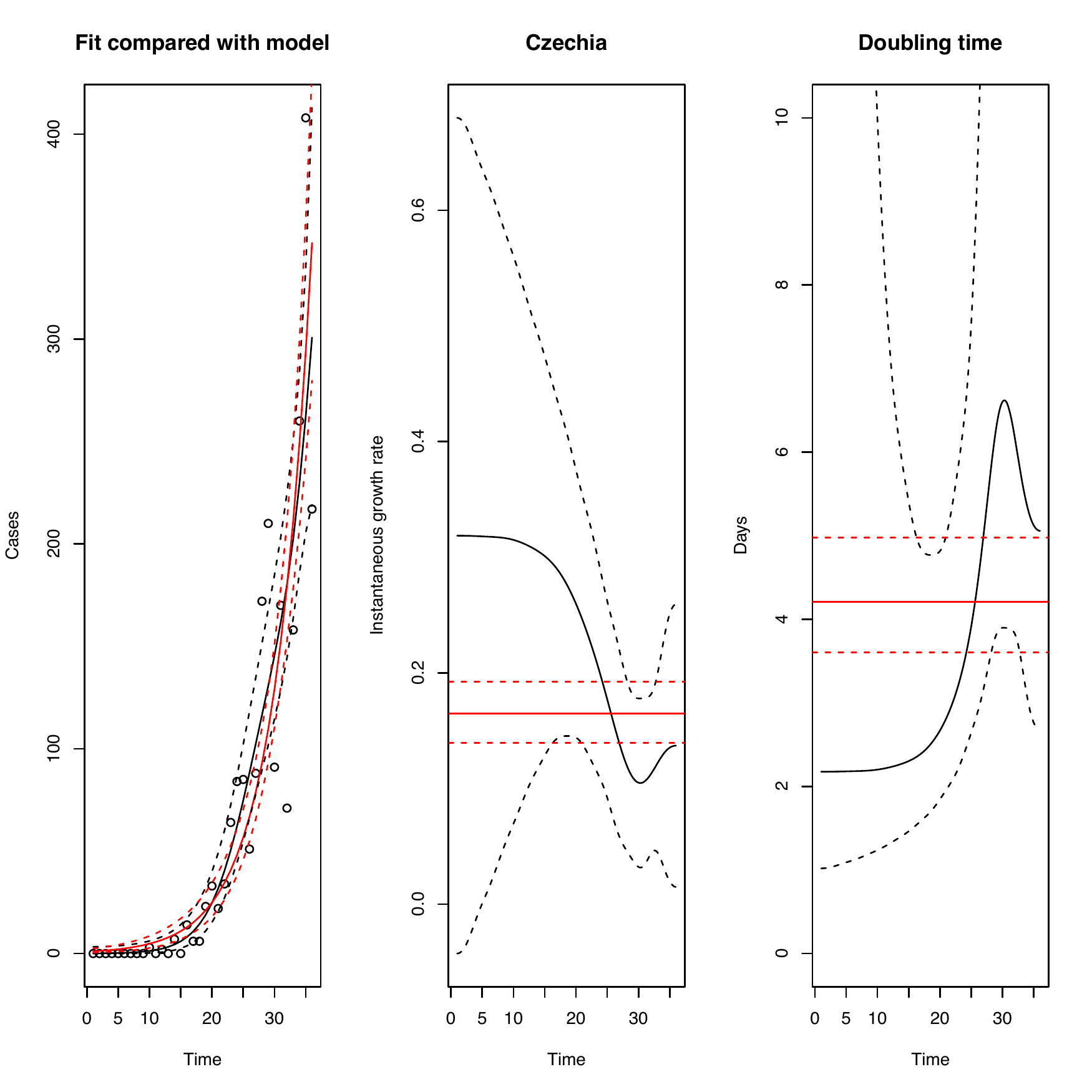}
\end{subfigure}
\begin{subfigure}{.28\textwidth}
    \centering
    \includegraphics[width=0.95\textwidth]{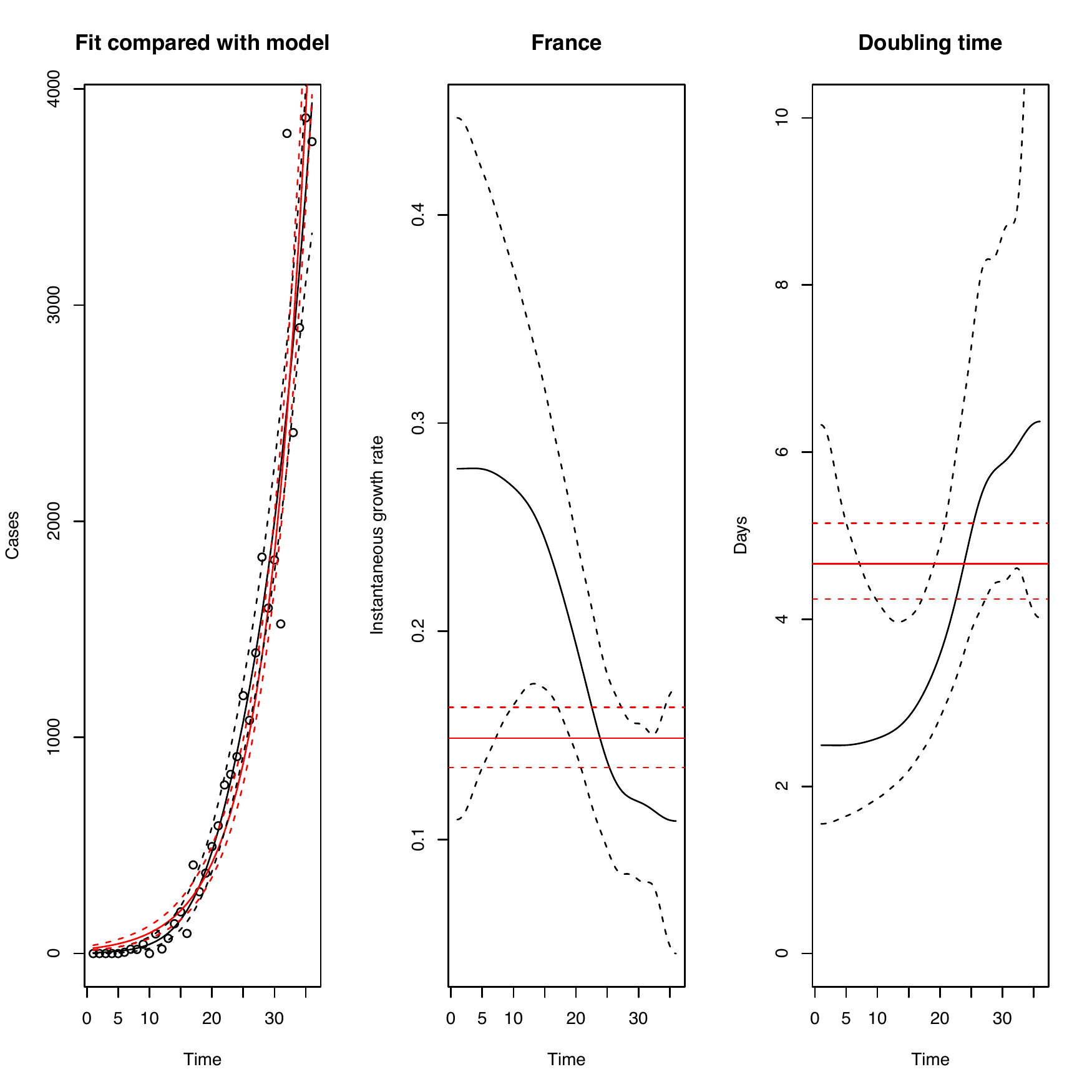}
\end{subfigure}%
\begin{subfigure}{.28\textwidth}
    \centering
    \includegraphics[width=0.95\textwidth]{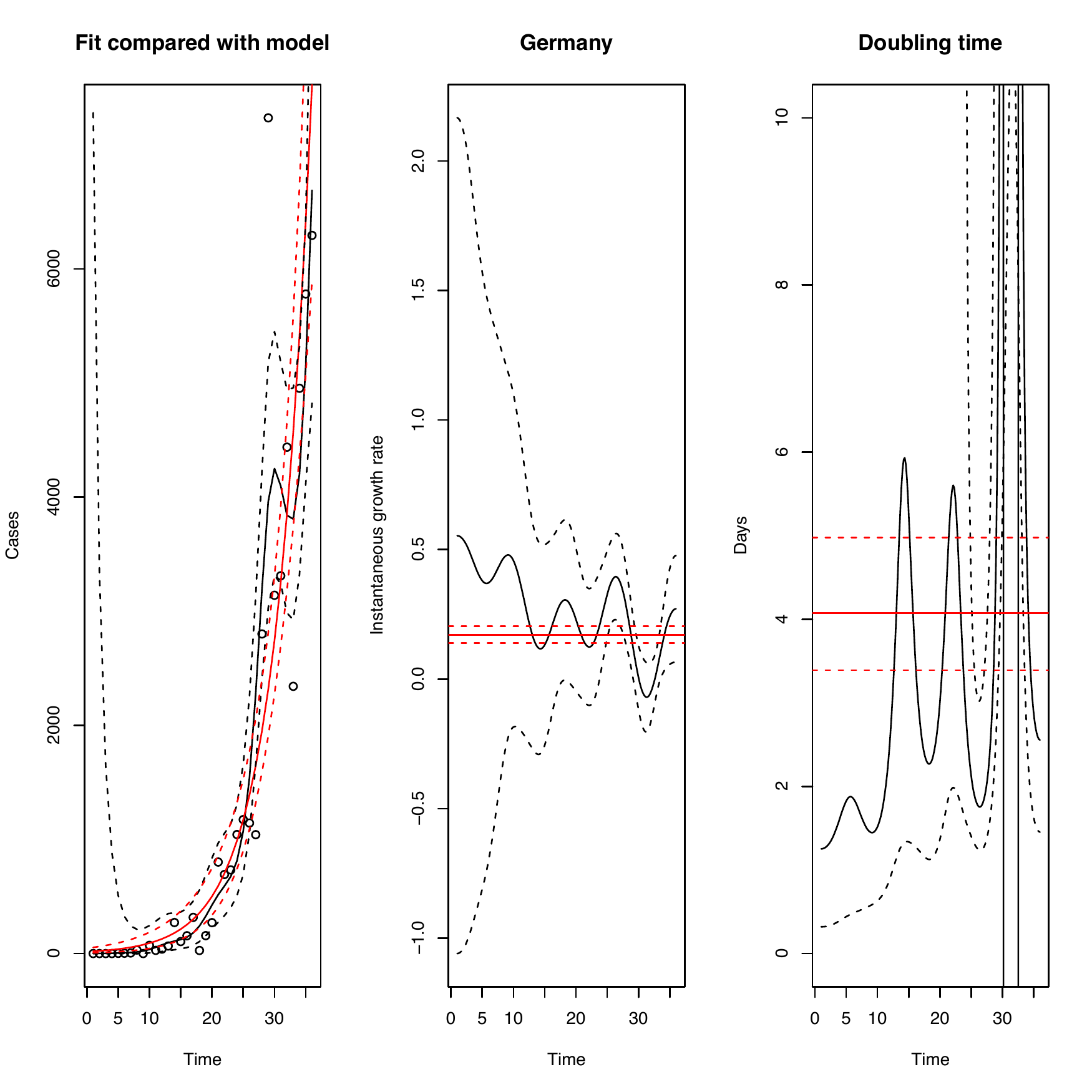}
\end{subfigure}%
\begin{subfigure}{.28\textwidth}
    \centering
    \includegraphics[width=0.95\textwidth]{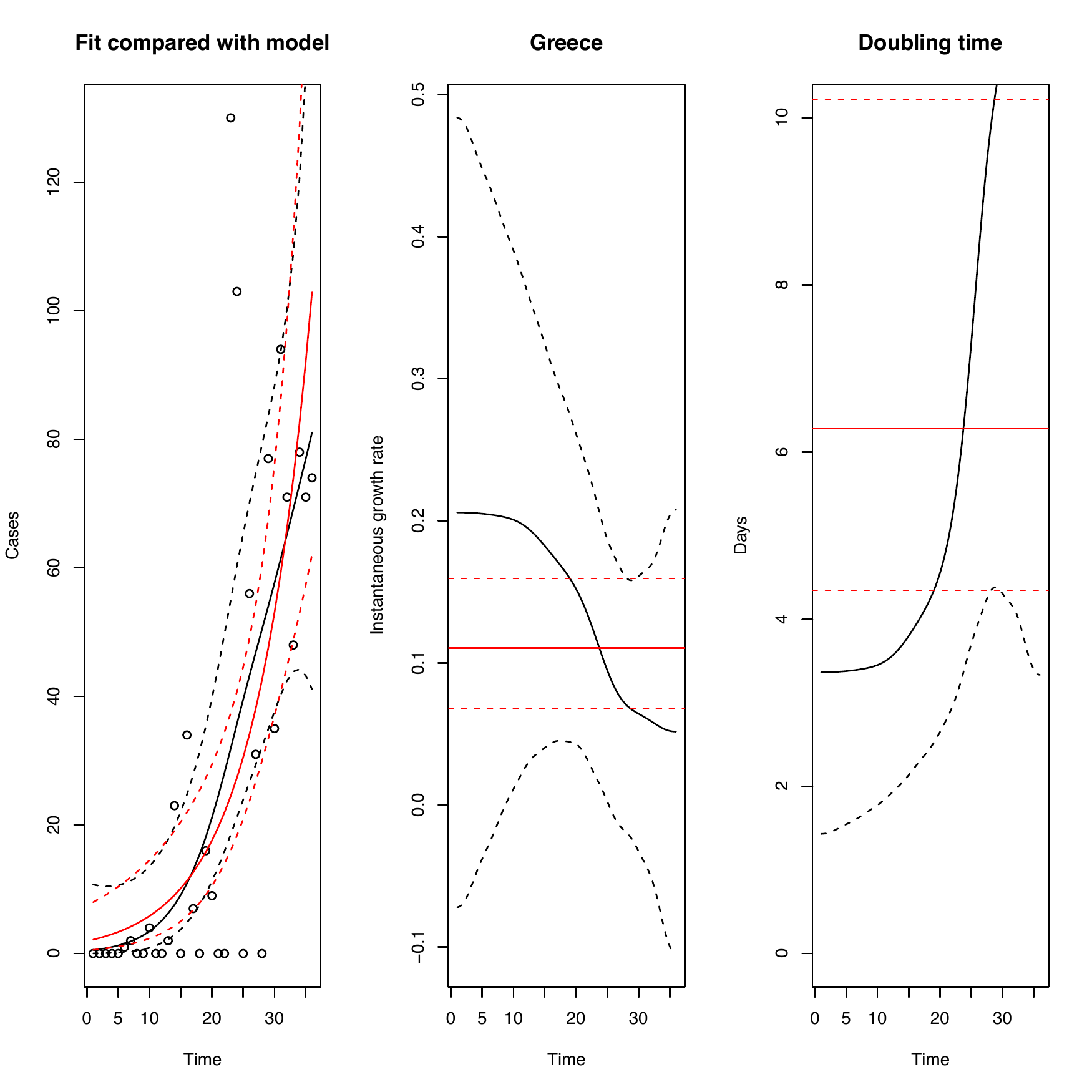}
\end{subfigure}
\begin{subfigure}{.28\textwidth}
    \centering
    \includegraphics[width=0.95\textwidth]{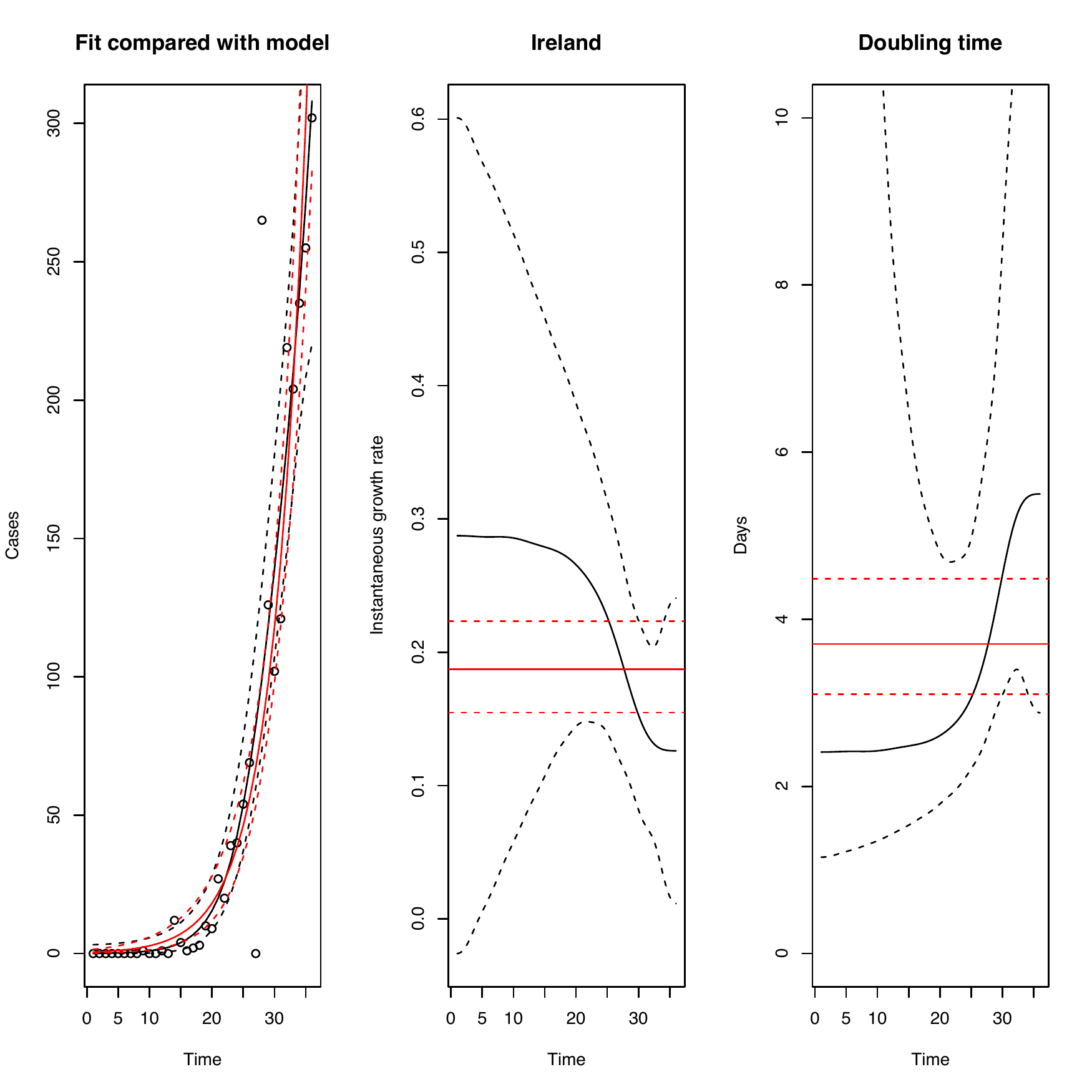}
\end{subfigure}%
\begin{subfigure}{.28\textwidth}
    \centering
    \includegraphics[width=0.95\textwidth]{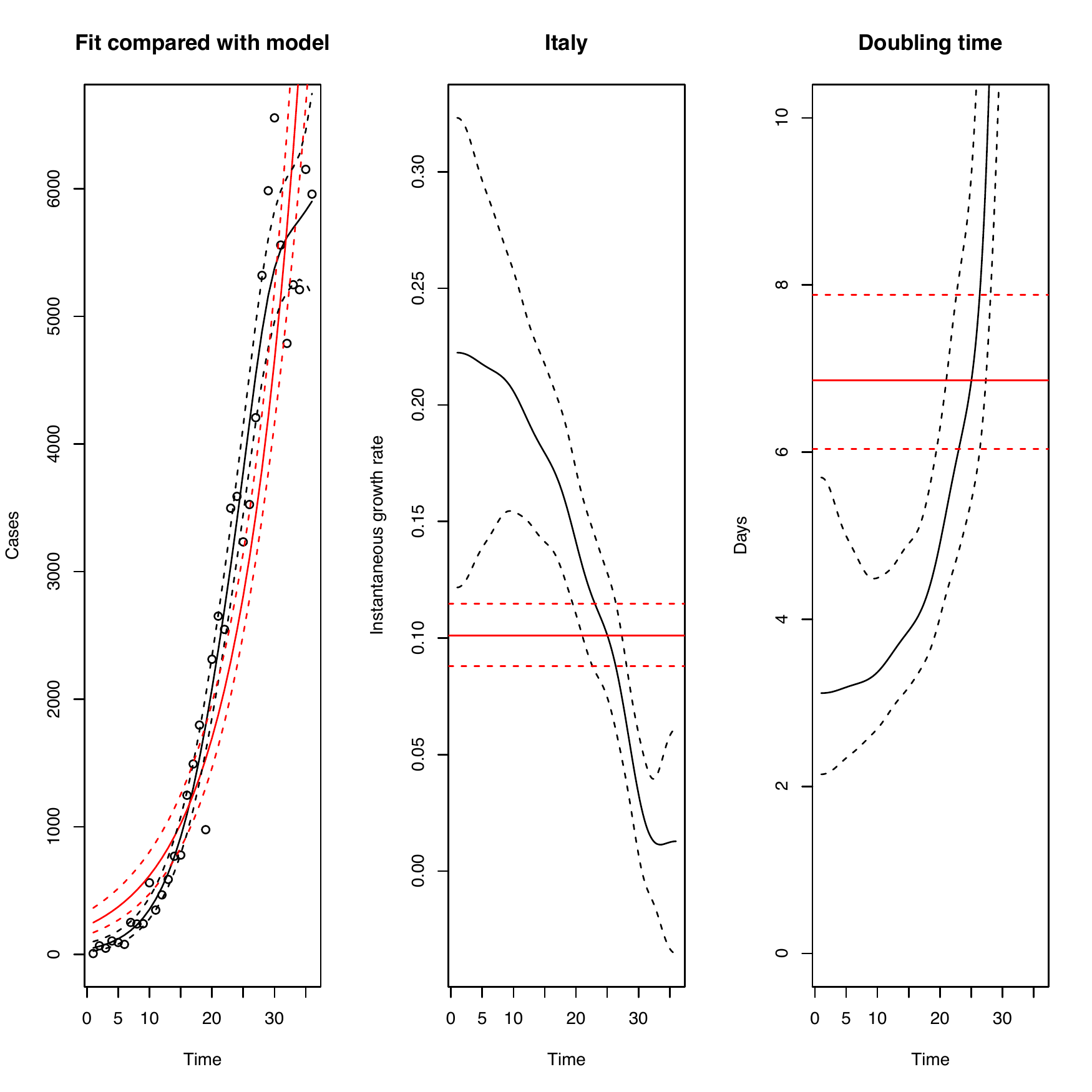}
\end{subfigure}%
\begin{subfigure}{.28\textwidth}
    \centering
    \includegraphics[width=0.95\textwidth]{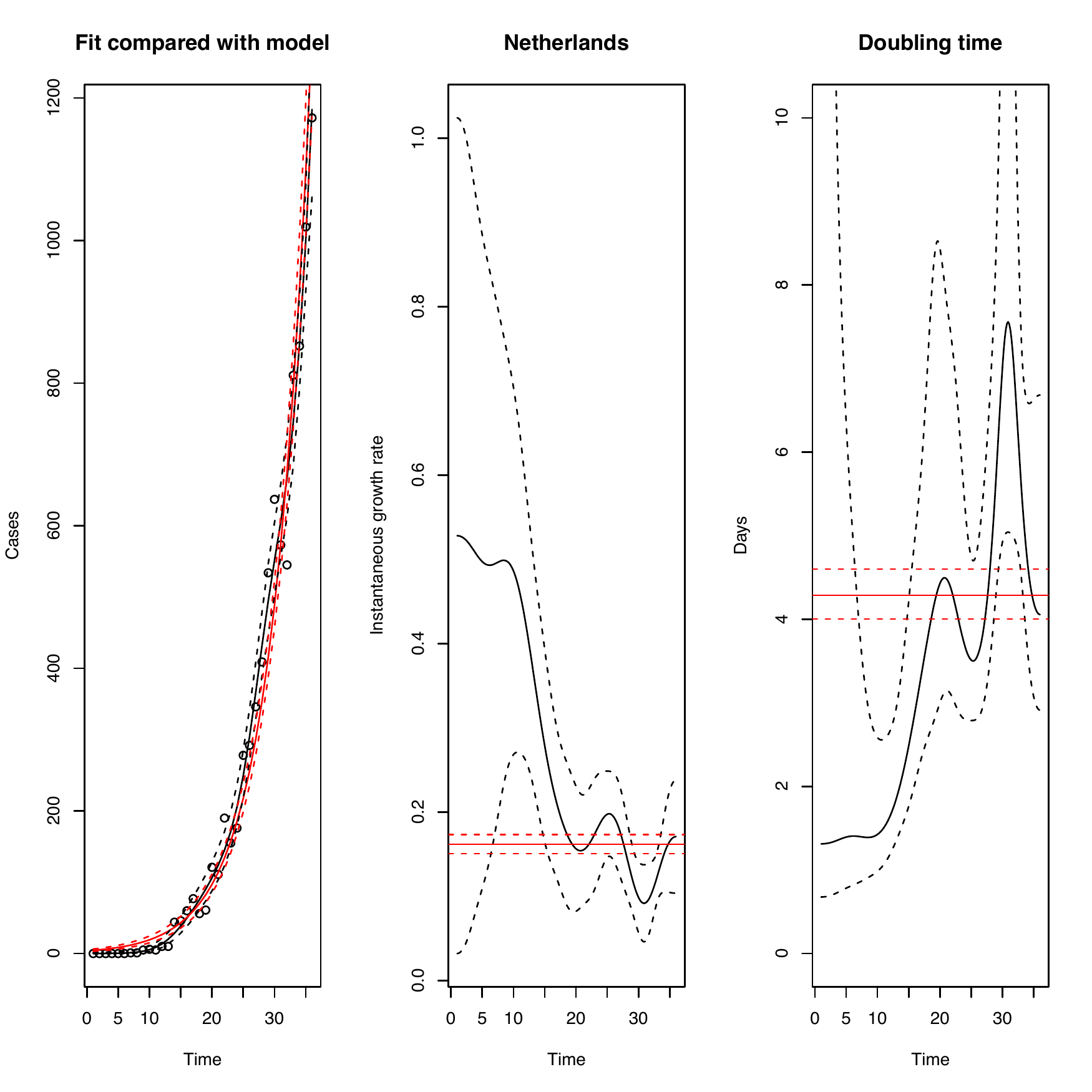}
\end{subfigure}
\begin{subfigure}{.28\textwidth}
    \centering
    \includegraphics[width=0.95\textwidth]{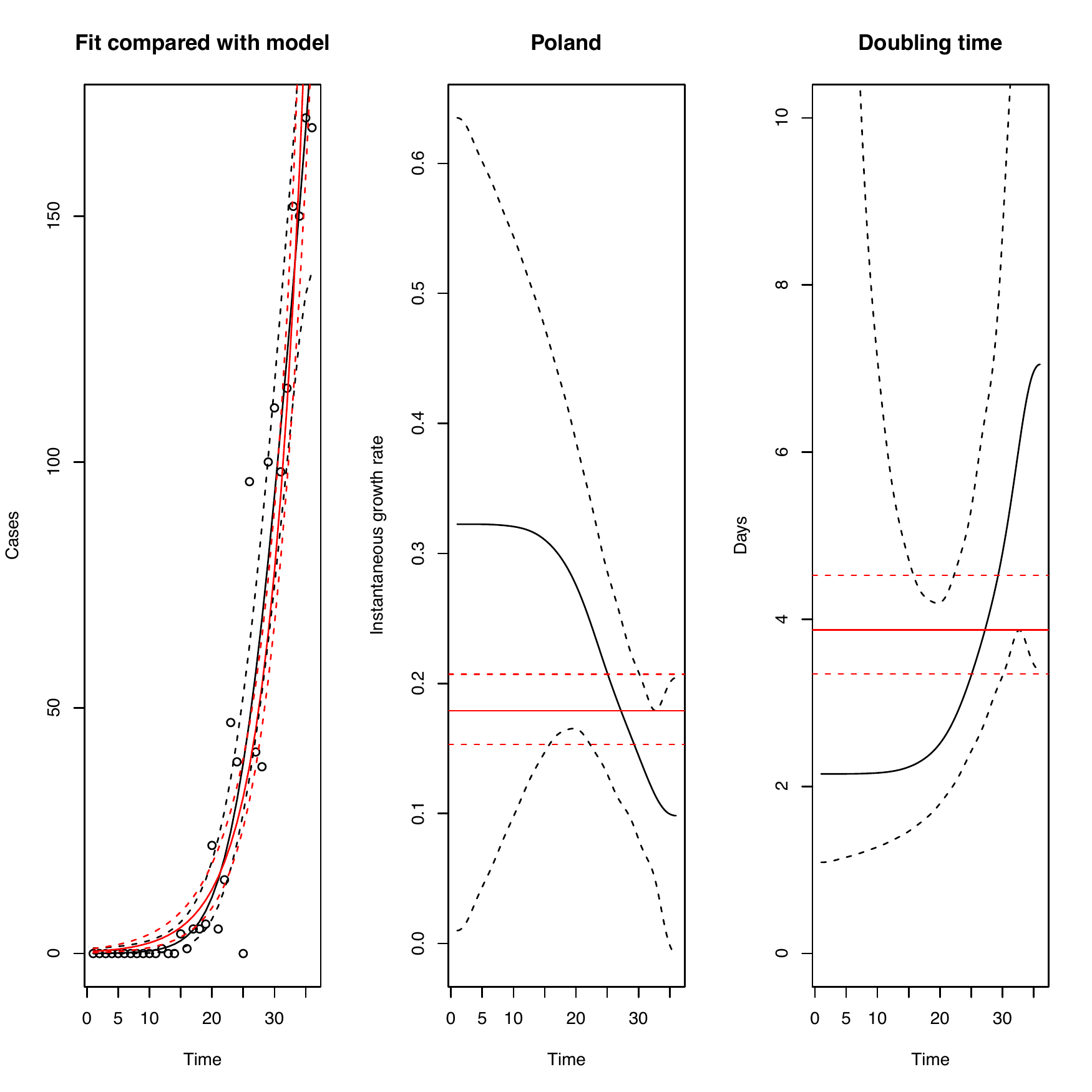}
\end{subfigure}%
\begin{subfigure}{.28\textwidth}
    \centering
    \includegraphics[width=0.95\textwidth]{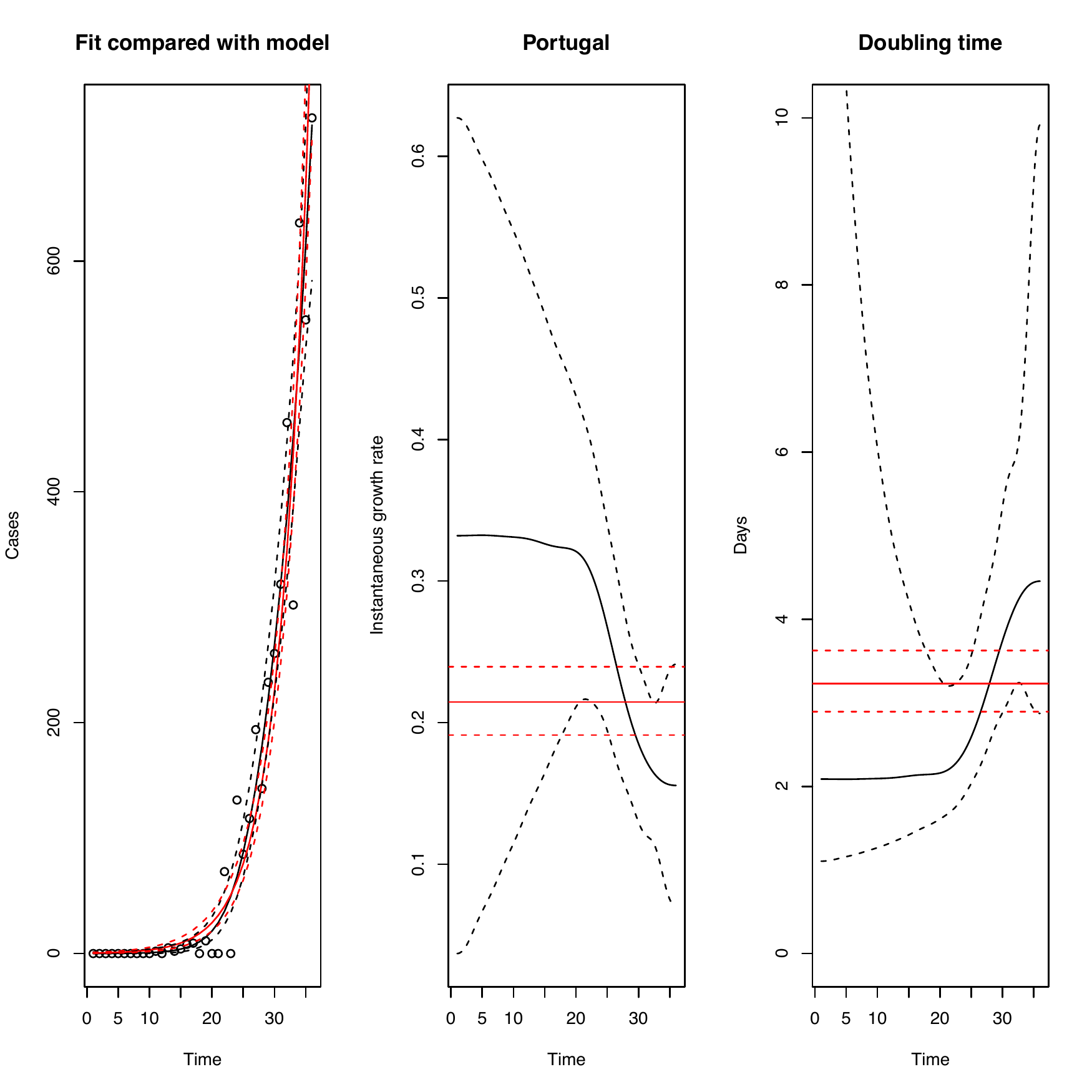}
\end{subfigure}%
\begin{subfigure}{.28\textwidth}
    \centering
    \includegraphics[width=0.95\textwidth]{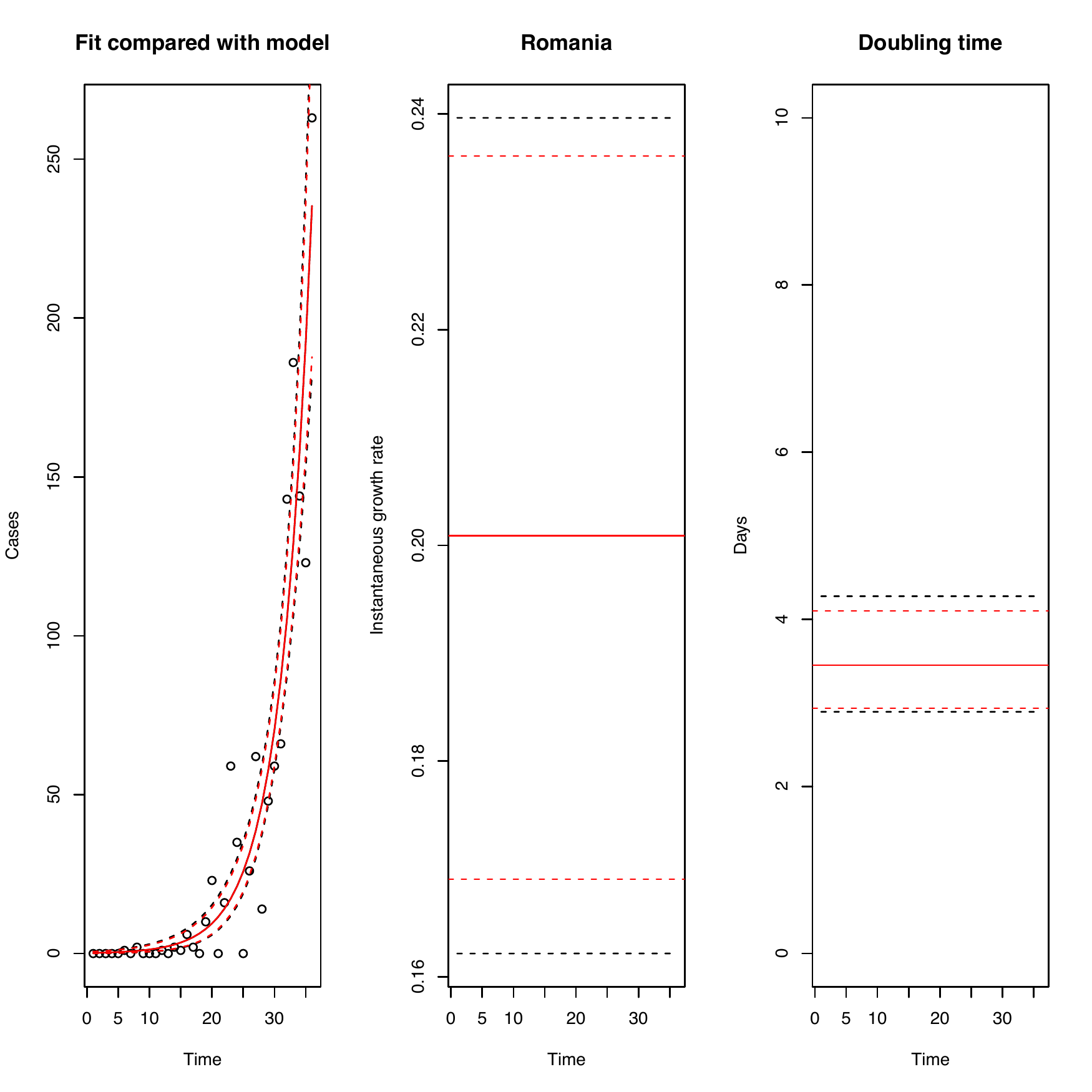}
\end{subfigure}
\begin{subfigure}{.28\textwidth}
    \centering
    \includegraphics[width=0.95\textwidth]{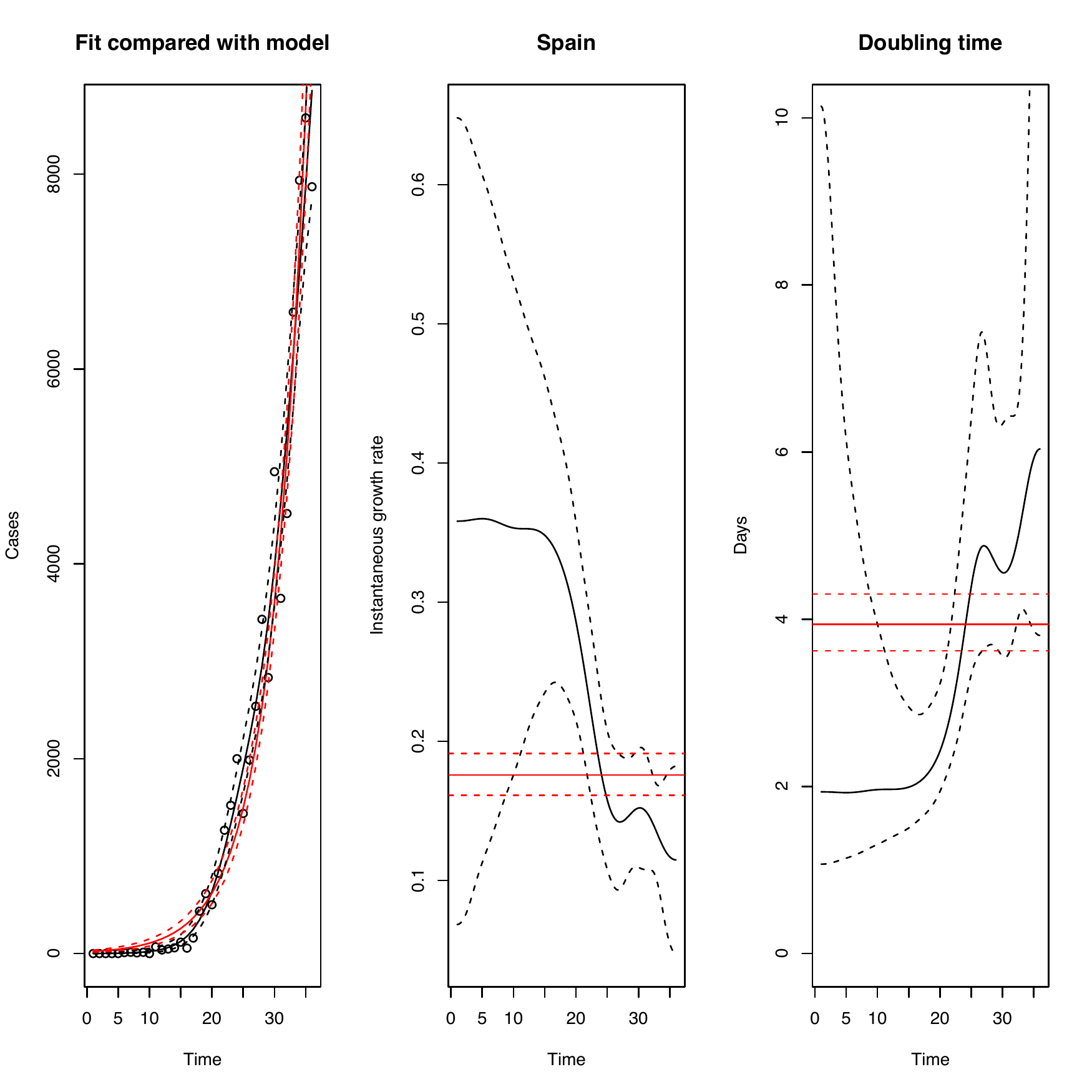}
\end{subfigure}%
\begin{subfigure}{.28\textwidth}
    \centering
    \includegraphics[width=0.95\textwidth]{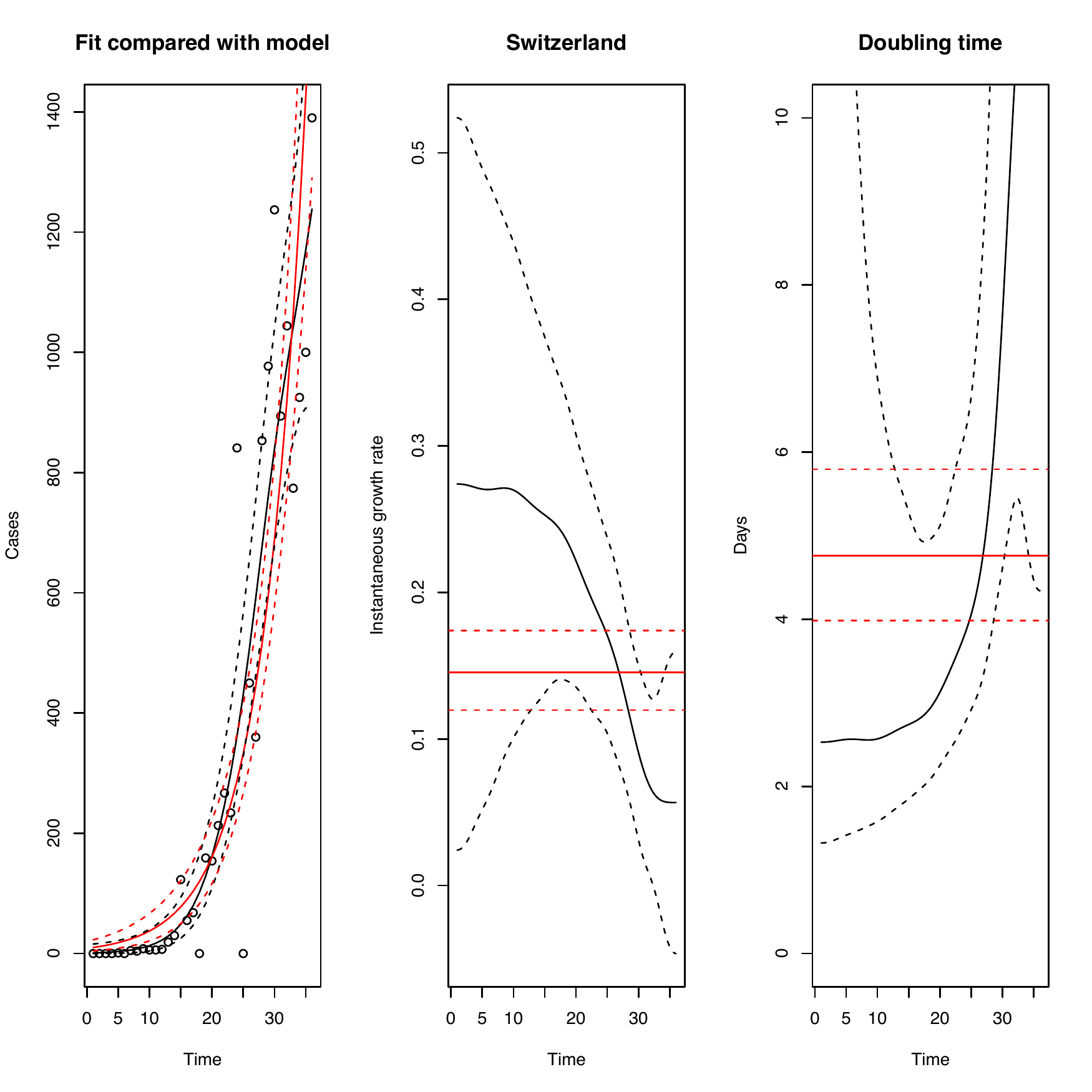}
\end{subfigure}%
\begin{subfigure}{.28\textwidth}
    \centering
    \includegraphics[width=0.95\textwidth]{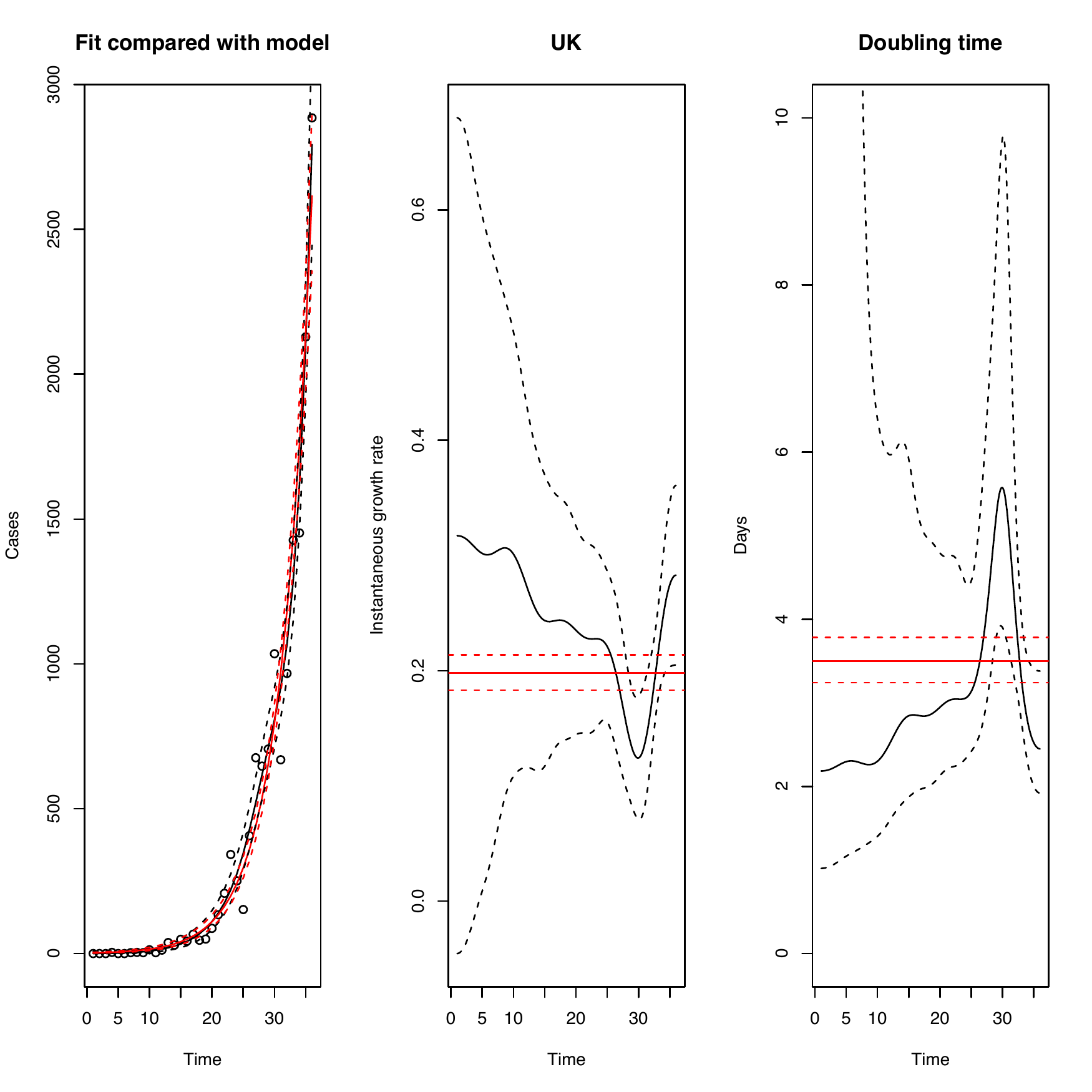}
\end{subfigure}
    \caption{GAM and GLM model fit for multiple countries. Left panel: model fit and data; middle panel: instantaneous growth rate from the GAM (black) and the (constant) growth rate from the GLM (red) with 95\% CI; right panel: doubling time from GAM (black) and GLM (red). The $x$-axis gives time since 21 February 2020.}
    \label{Fig:MajordoublingIncidence}
\end{figure}

\begin{figure}[h!]
    \centering
    \includegraphics[width=0.9\textwidth]{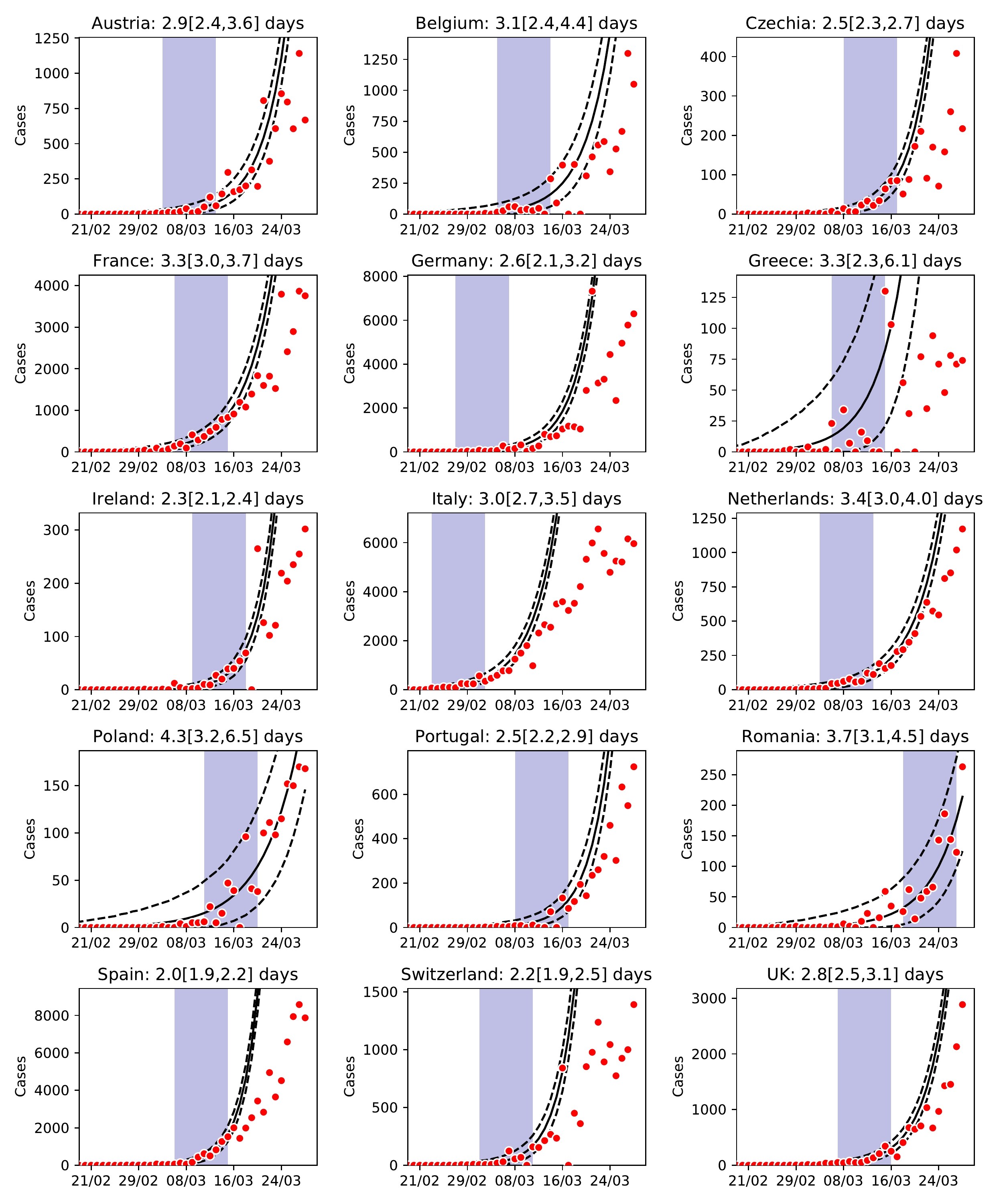}
    \caption{Linear version of Figure 2 from the main paper.}
    \label{Fig:linearversion}
\end{figure}

Each page of Figure \ref{Fig:MajordoublingIncidence} shows the GAM compared to a simple GLM
with $\theta = 1$ and $\mathcal{T}$ taken to be all of the
data range, in contrast to the results in the main paper and Figure \ref{Fig:linearversion}, where
we fit $\theta$ and let $\mathcal{T}$ correspond to the first nine days of the local epidemic after the cumulative number of cases has reached 25 (the only exceptions are the UK and Romania, where fitting started an additional 9 days later to reflect the local situation). While the simple GLM method is clearly inadequate if the fit is not restricted to a window where exponential growth appears reasonable, it is shown for comparison in these plots. The left panel shows the output of the model fit and the data, the middle panel the instantaneous growth rate from GAM (black) and the growth rate from GLM (red) with 95\% CI, and the right most panel shows these growth rates converted to doubling times. Of the fifteen countries, two (Belgium and Romania) have equivalent fits from GAM as from GLM and show constant growth rates over the time period. 
The Czech Republic, Greece, Ireland and Poland have the central GLM result within the 95\% CI of the GAM suggesting a constant growth rate is a plausible explanation of the data reported. Austria, France, Italy, Portugal, Spain, Switzerland show a fairly smooth transition from short to longer doubling times. Germany, Netherlands and UK show more oscillatory behaviour in doubling times.

\begin{figure}[h!]
    \centering
    \includegraphics[width=0.9\textwidth]{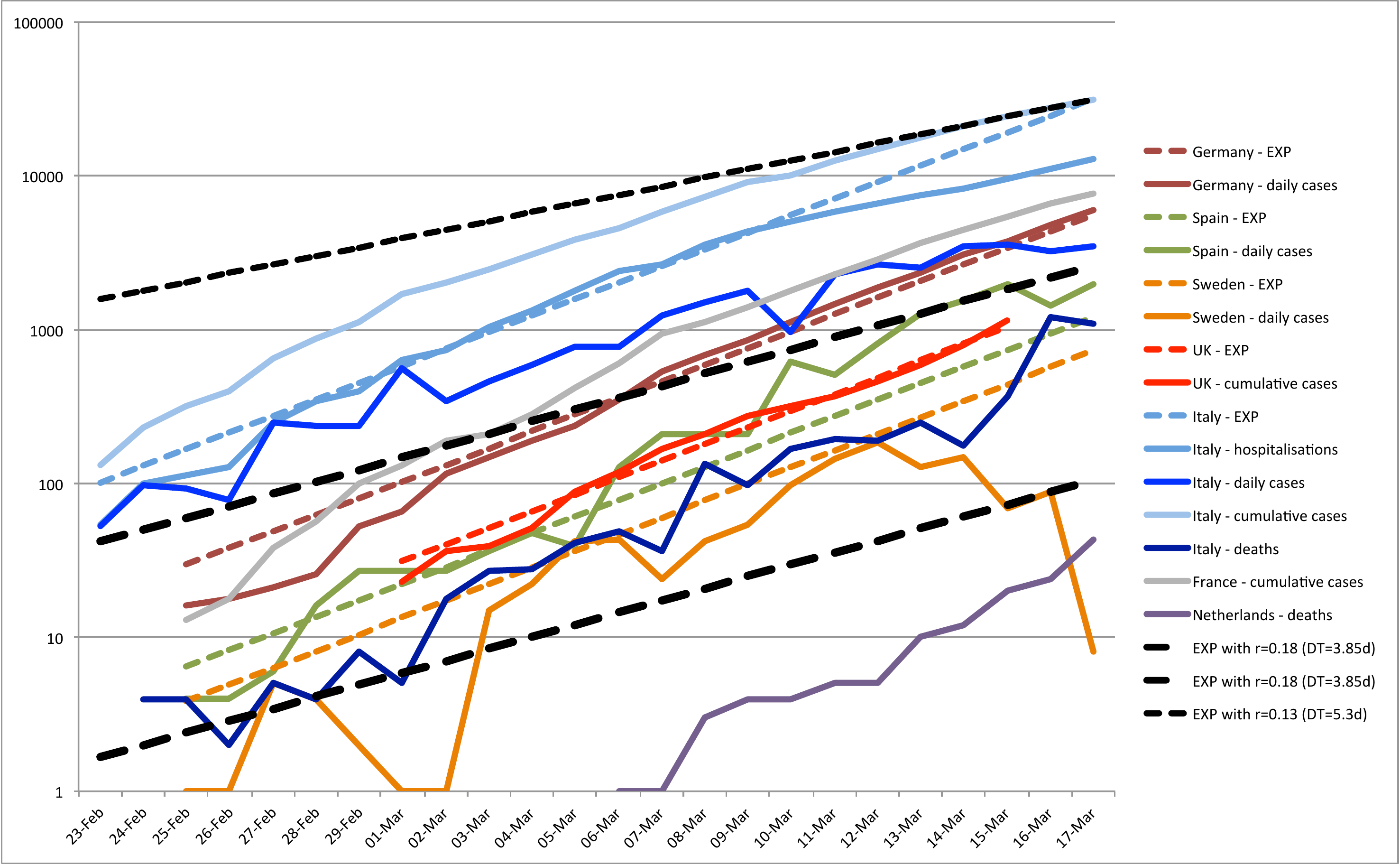}
    \caption{Growth rates in different EU countries and visual comparison with simple exponential growth.
A `mixed bag' of numbers of daily new confirmed cases, cumulative confirmed cases, hospitalisations and deaths, all showing similar fast growth. Visible exceptions are: Sweden, which has changed testing regime; Italy, where growth in new confirmed and cumulative confirmed cases appears to be slowing down from beginning of March (quarantining infected towns from 22-23 Feb) and growth in hospitalisations around a week later (deaths not slowing down yet); deaths in Netherlands and new cases in Spain, which are growing even faster than the rest. Coloured dashed lines are pure exponential growth with $r = 0.25$ day$^{-1}$ (doubling time of 2.77 days, in the same ballpark as UK and Italy). No statistical fit is performed: lines’ slope and intercept are purely eyeballed. Black thick dashed lines show growth at $r = 0.18$ day$^{-1}$ (doubling time of 3.85 days).
Black thin dotted line shows growth at $r = 0.13$ day$^{-1}$ (doubling time of 5.33 days). Notice how this is roughly the growth in cases (new and cumulative) and hospitalisations in Italy around 17 March.}
    \label{Fig:growthrates}
\end{figure}

As a further test for the robustness of our results, we simply visually assess the growth rates of the epidemic in a set of countries by plotting data on a log scale and compare the observed slopes with pure exponential trends (Figure \ref{Fig:growthrates}). To avoid relying on confirmed cases only, we plot a `mixed bag' of cumulative cases, daily new confirmations, hospitalisations and deaths. We also plot exponential trends with a lower growth rate ($r=0.18$ and $r=0.13$) to aid the visual distinction in slopes between different exponential growths and the visual assessment of the potential effects of interventions in Italy. \\

\noindent\underline{Estimating delay distributions}\\
\indent Delay distributions describe the time  delay between two events. To understand how long until the impact of an intervention may be observed, we need to understand the delay between infection and symptom onset (the incubation period), and the delay from onset to hospitalisation. A difficulty with estimating delay distributions during an outbreak is that events are only observed if they occur before the final sampling date. Since delay distributions depend on the time between two events, if the first event occurs near to the end of the sampling window, it will only be observed if the delay to the second event is short. This causes an over-expression of short delays towards the end of the sampling window, which is exacerbated by the exponential growth of the epidemic. Therefore, we need to account for this growth and truncation within our model.

To fit the data we use maximum likelihood estimation. However, we do not observe the delay directly, instead observing the timing of the two events. Therefore, we need to construct a likelihood function for observing these events. Following [\textit{49}], we construct the conditional density function for observing the second event given the time of the first event and given that the second event occurs before date $T$. That is, we are interested in the conditional density function 
\begin{equation}
L(X_2 \in x_2|X_1 \in x_1,X_2\leq T)=\frac{L(X_2 \in x_2,X_1 \in x_1,X_2\leq T)}{L(X_1 \in x_1,X_2\leq T)},
\end{equation} where $x_1$ and $x_2$ can be exactly observed or interval censored.

The delay from onset to hospitalisation for the UK is estimated using FF100 data provided by Public Health England, which contains data on the first few hundred infected individuals in the UK. This data incorporated the time of symptom onset and time of hospitalisation. There were some cases who were hospitalised before their onset date. These cases have been removed from the data set, since they do not provide insight into the delay. Additionally, some cases have no symptom onset, so these have also been removed from the data. For cases where symptom onset and hospitalisation occur on the same day, we add half a day to the hospitalisation day, since the delay is unlikely to be instantaneous. After tidying the data, this left 106 cases from which to infer the onset to hospitalisation delay. The dates in the line list are recorded exactly, so the likelihood function becomes
\begin{equation}
L(H=h|O=o,H\leq T)=\frac{g(o)f(h-o)}{\int_0^{T-o}g(o)f(x)\mathrm{d}x}
=\frac{f(h-o)}{\int_0^{T-o}f(x)\mathrm{d}x},
\label{eqn:trunc1}
\end{equation}
where $f$ is the density of the onset to hospitalisation delay and $g$ is the density of the onset time. Using this truncation corrected method and a gamma distribution to fit the delay distribution, we get a mean delay of 5.14 with standard deviation 4.20. Unfortunately, we cannot share the FF100 data. To compare different regions, we also use data from Hong Kong and Singapore to estimate the local onset to hospitalisation delays. This data is taken from an open access line-list [\textit{33}], and the filtered data sets used are provided in the supplementary material. Using the method above, for Hong Kong the mean delay is 4.41 days, with standard deviation 4.63, and for Singapore the mean delay is 2.62 with standard deviation 2.38. 

For the incubation period, we use data from Wuhan during the early stages of the outbreak. This data was extracted from an open access line-list [\textit{33}], containing dates when individuals were in Wuhan and when they developed symptoms (among other information). Since this data is from the early stages of the epidemic, the majority of cases were in Wuhan. Therefore, it is likely that these individuals were infected in Wuhan, so the time spent in Wuhan provides a potential exposure window during which infection occurred. For individuals with symptom onset date before leaving Wuhan or the same day they left Wuhan, the upper bound on the exposure window was adjusted to half a day before symptom onset. Using the data as of 21/02/2020, we have 162 cases from which to infer the incubation period. This infection date is interval censored, so we obtain the likelihood function

\begin{equation}
L(O=o|a<I<b,O\leq T)=\frac{\int_a^bg(i)f(o-i)\mathrm{d}i}{\int_a^b\int_0^{T-i}g(i)f(x)\mathrm{d}x\mathrm{d}i},
\label{eqn:trunc2}
\end{equation} 
where $f$ is the incubation period density function and $g$ is the density of the infection date. We assume $g$ is proportional to force of infection of the outbreak, which is assumed to follow exponential growth with rate parameter $0.25$ day$^{-1}$. Using a gamma distribution to describe the incubation period, we get a mean incubation of 4.84 days with standard deviation 2.79. The data for the incubation period is also included as supplementary material, along with MATLAB code to perform the maximum likelihood estimation.\\

\noindent\underline{Estimation of $R_0$}\\
\indent The relationship between the growth rate $r$ and the basic reproduction number $R_0$ in a simple homogeneously mixing model is provided by the Lotka-Euler equation:
\begin{equation}
\frac{1}{R_0} = \int_0^\infty {\omega(\tau) \mathrm{e}^{-r\tau}\mathrm{d}\tau},
    \label{eq:LotkaEuler}
\end{equation}
where $\tau$ represents the time since the infection of an individual and  $\omega(\tau)$ is the infectious contact interval distribution, defined as the probability density function (pdf) of the times (since infection) at which an infectious contact is made. An infectious contact is a contact that results in an infection if the contactee is susceptible, and early on in the epidemic any randomly selected contactee is almost surely susceptible.

\begin{figure}[h!]
    \centering
    \includegraphics[width=0.9\textwidth]{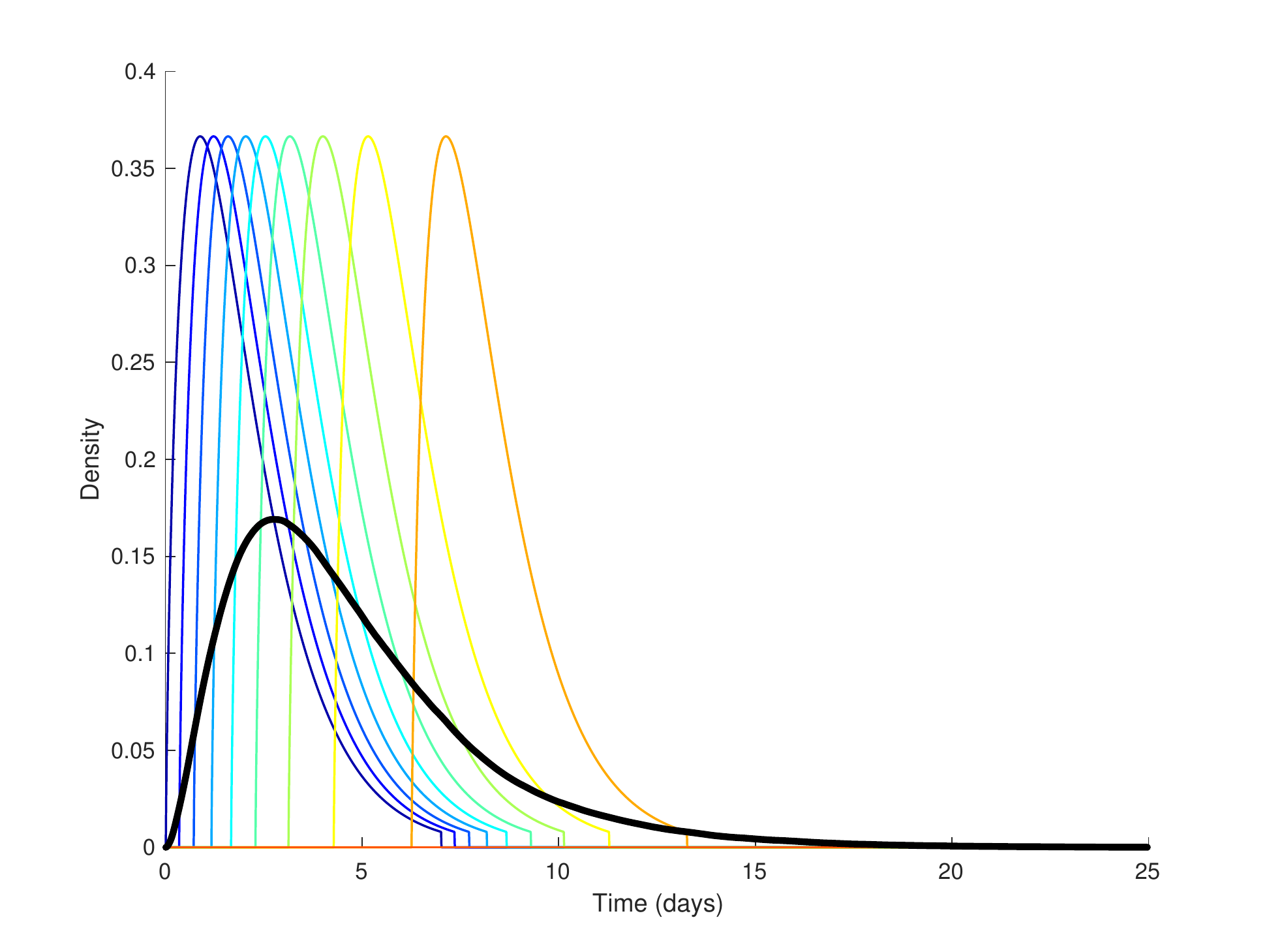}
    \caption{Random variability in the generation time distribution (thin coloured lines) and time-point average (thick black line).}
    \label{Fig:randomTVI}
\end{figure}

Equation \eqref{eq:LotkaEuler} assumes all individuals have the same infectious contact interval distribution. However, if we assume random variability between individuals, there will be a set $\mathcal{S}$ of curves $\Omega(\tau)$. However, equation \eqref{eq:LotkaEuler} still applies, with $\omega(\tau)$ being the time-point average of all curves in $\mathcal{S}$ [\textit{50}]; see Figure \ref{Fig:randomTVI}.

The generation time $T_g$ is defined as the mean of the infectious contact interval distribution $\omega$:
\begin{equation}
T_g = \int_0^\infty {\tau \omega(\tau)\mathrm{d}\tau}.
    \label{eq:Tg}
\end{equation}{}
The same definition extends to a random infectivity profiles of which $\omega$ is the time-point average.

For the incubation period we use our estimates from Table 1 (mean 4.84, standard deviation 2.79), which are anyway similar to those estimated by others. However, information about any form of pre-symptomatic transmission is hard to obtain but crucial for $R_0$ estimates [\textit{34-37}]. Furthermore, there is also limited information concerning how infectivity changes over time. Therefore, Table \ref{tab:R0} reports the estimates we obtain assuming the infectious period starts at the onset of symptoms, one, two or three days earlier, and assuming a Gamma-shaped infectivity with mean 2 or 3 days. In both cases, the standard deviation is assumed to be 1.5 and the infectivity is truncated after 7 days (see Figure \ref{Fig:randomTVI}).

We conclude that the estimates of $R_0$ are highly sensitive to small variations in quantities that are poorly supported by available data, but that for a growth rate of 0.25 day$^{-1}$, close to what is observed in Italy and the UK, are also generally larger, and possibly much larger, than official estimates [\textit{3}, \textit{12-15}]. Smaller values in this range are associated with significant amounts of pre-symptomatic transmission [\textit{34}], leading to a generation time for example compatible with some of the shortest estimates of the serial interval seen in the literature [\textit{51}], and with a front-loaded infectivity curve (mean 2, rather than 3).

We tested further assumptions. A simple SEIR model, with exponentially distributed incubation and infectious periods (with the same means as above but constant infectivity) leads to much smaller values of $R_0$ than our estimates, as it favours really short incubation periods (Table S1B, left). Estimates, instead, do not change significantly if high variability in total infectiousness between individuals, in line with what observed for SARS, is assumed (Table S1B, right) or if 50\% of cases are assumed to be fully asymptomatic and transmit at half the rate as those with symptoms (not shown).

These simple estimates are obtained under the assumption of mass-action mixing. The explicit presence of a social structure (e.g.\ age-stratification, household/network structure, etc.), which in principle could affect them, is likely negligible in such a high $R_0$ and growth rate regime [\textit{52}]. The effect of the social structure on transmission is expected to grow in importance (especially the household structure, since isolation and quarantine facilitate within-household transmission) the closer $R_0$ is to 1.

\newpage

\newpage

\newpage

\newpage

\newpage
\begin{table}[H]
    \centering
    \includegraphics[width=0.9\textwidth]{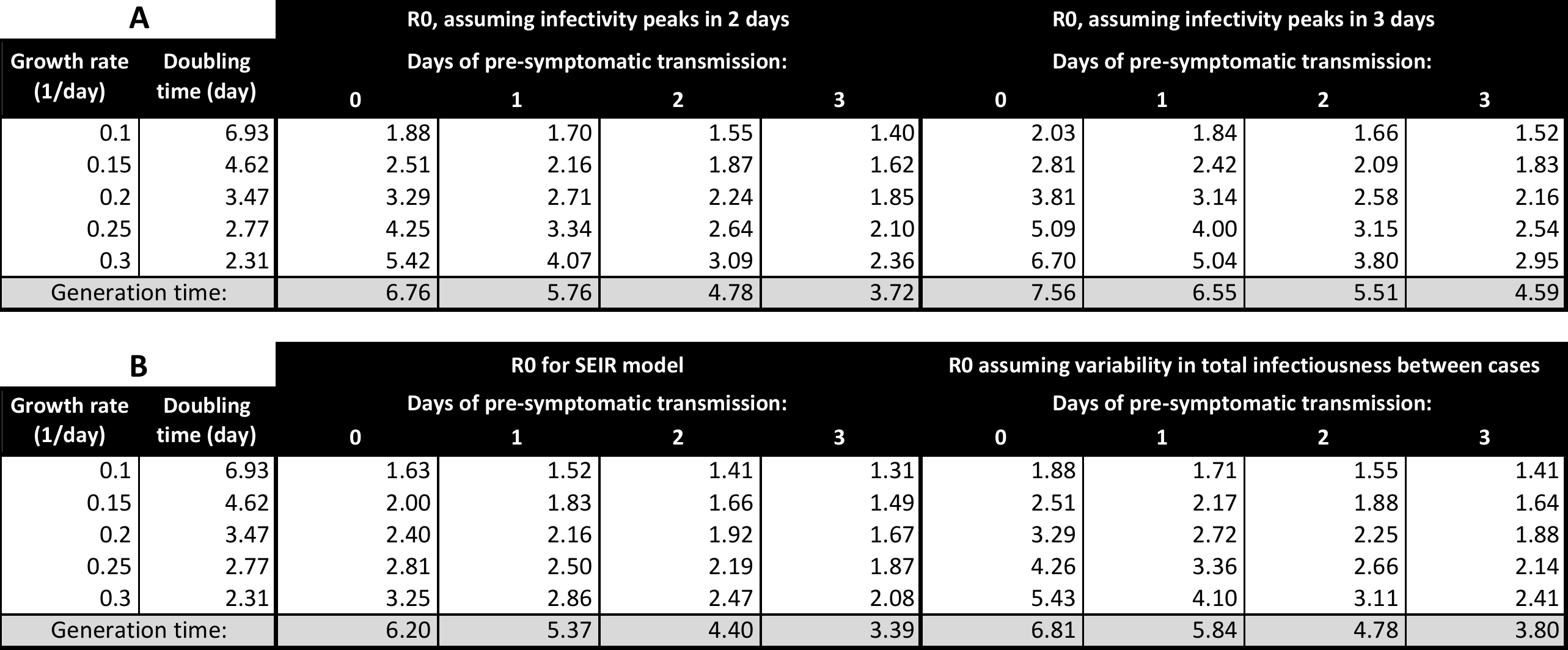}
    \caption{Values of $R_0$ derived from different growth rates and different modelling assumptions. A) Gamma-distributed latent period with estimates from Table 1, and Gamma-shaped infectivity profile with mean 2 (left) and 3 (right) and standard deviation 1.5; B) SEIR model (left) and same model as in A) but assuming total infectiousness is randomly drawn from a Gamma-distribution with mean 1 and standard deviation $1/\sqrt{k}$, with $k=0.25$.}
    \label{tab:R0}
\end{table}



\end{document}